# Distributed computing of Seismic Imaging Algorithms

*Masnida Emami* [1,2], *Ali Setayesh*[1], *Nasrin Jaberi* [3]

[1] *School of Science, Tarbiat Moaleem University, Karaj, Iran*
[2] *University College of Engineering and Technology Malaysia, Kuantan*
[3] *Payame-noor University, Najafabad*
Nasrinjaberi60@yahoo.com

## 1. Abstract

The primary use of technical computing in the oil and gas industries is for seismic imaging of the earth's subsurface, driven by the business need for making well-informed drilling decisions during petroleum exploration and production. Since each oil/gas well in exploration areas costs several tens of millions of dollars, producing high-quality seismic images in a reasonable time can significantly reduce the risk of drilling a "dry hole". Similarly, these images are important as they can improve the position of wells in a billion-dollar producing oil field. However seismic imaging is very data- and compute-intensive which needs to process terabytes of data and require Gflop-years of computation (using "flop" to mean floating point operation per second). Due to the data/computing intensive nature of seismic imaging, parallel computing are used to process data to reduce the time compilation.

With introducing of Cloud computing, MapReduce programming model has been attracted a lot of attention in parallel and distributed systems [1, 2] to execute massive processing algorithms such as Bioinformatics[3] , Astronomy[4], Geology[5] and  so on. In this report, we will investigate and discuss current approaches to fit seismic algorithms to MapReduce programming model.

## 2. Seismic imaging

The input seismic data is a collection of tens or hundreds of thousands of seismic experiments. Each experiment consists of a single source of sound waves and a collection of hundreds of recording instruments, all located at the surface of the earth. Sound waves propagate from an impulsive source (called a "shot") down into the earth, reflect off boundaries between regions in the subsurface and propagate back to the surface. Each recording instrument (Geo-phones) listens to these returning "echos" for ten or more seconds at a sampling rate of a few milliseconds. Consequently, the data for each of these receiver recordings (called a seismic "trace") is of order ten kB in size. Each experiment yields a collection of traces, called a "shot record", which contains information on reflecting surfaces in the neighborhood of the source location. Other nearby shots will contain reflections off many of the same surfaces, though with somewhat different propagation paths, thus providing for redundancy of information on the position of reflectors. An entire seismic survey, with shots located over an area of hundreds or thousands of km$^2$, consists of tens of millions of seismic traces, for a total dataset size of order a terabyte (TB).

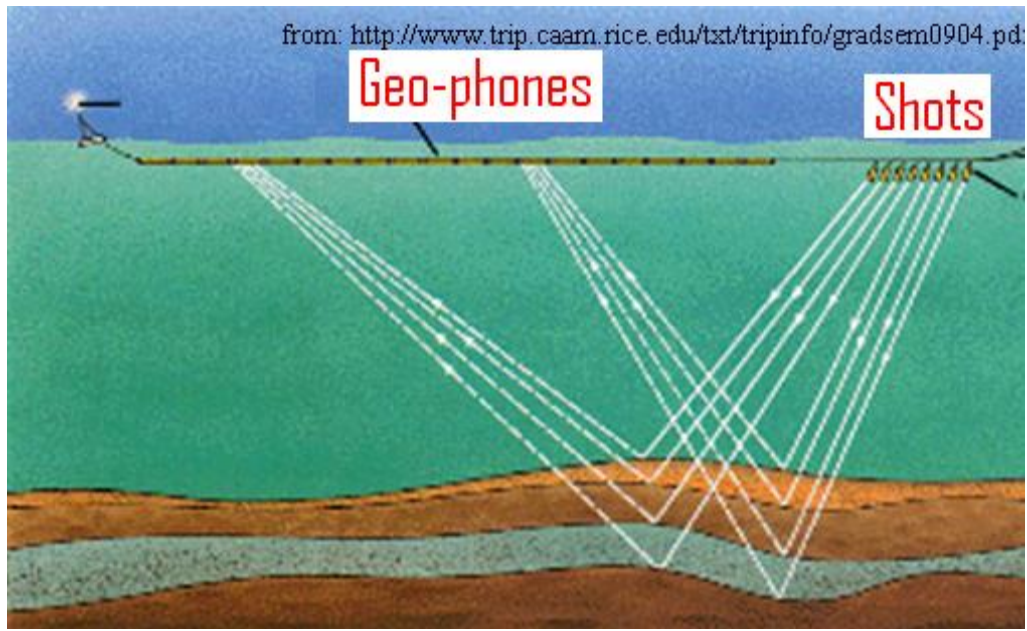

Seismic imaging, the process of determining an accurate image of the subsurface of the earth from all these "echos", is an inverse problem which is solved in an approximate and iterative fashion. Each iteration uses an estimate of the regional acoustic velocity to back propagate (or "migrate") the surface seismic data down into the earth, determining approximately where the reflections took place [6]. Because of the redundancy in the seismic data, this migration process results in a set of redundant images of the boundaries between subsurface layers. Differences in the positioning of the same subsurface reflector in different images provide a measure of the error of the model of the propagation velocity. If the error is significant, the velocity model can be corrected and another migration of the seismic data can be performed. When the error is small, the redundant images are added (or "stacked"), improving our signal-to-noise ratio, resulting in an image which can guide the drilling of oil wells.

## 3. Motivation to utilize High Performance Computing in Seismic imaging

The computationally expensive step in the iterative process of seismic imaging is the migration of the seismic data. While there are a number of methods for performing this process, varying in their accuracy and expense, the least approximate yet affordable approach commonly used today is based on the Kirchhoff integral. The seismic images are produced by performing integrals of the data over the source and receiver coordinates, $x_s$ and $x_r$, respectively. In this migration method, the redundant dimension is usually taken to be the distance between the source and receiver, called the "offset", $X = |x_s - x_r|$. The images are given by:

$$I(x, X) = \int dx_s dx_r W(x, x_s, x_r)\delta(X - |x_s - x_r|)D(x_s, x_r, t = T(x, x_s) + T(x, x_r)) \quad (1)$$

where $x$ is the subsurface image coordinate, $D(x_s, x_r, t)$ is the recorded seismic data, $W(x, x_s, x_r)$ is a weighting function accounting for the amplitude of propagation and $T(x, x_s)$ and $T(x, x_r)$ are the times for propagation in the model from the image point to the source and receiver, respectively. So each data sample is summed into the image at all the points where the measured time agrees with the total propagation time, and the reflectors in the image are built up by constructive interference of data. ([7] is a good reference for seismic migration methods.) The scale of computation involved becomes apparent when the dimensionality of the problem is considered. Since $x_s$ and $x_r$ are each 2-dimensional, then for each point $x$ in the 3-dimensional subsurface, Equation (1) requires that we perform a 4-dimensional integral over a complex surface in the 5-dimensional data set $D(x_s, x_r, t)$. The computational complexity is then approximately

$$f_K N_{xyz} N_{tr} \quad (2)$$

where $N_{xyz}$ is the number of image points in the subsurface, $N_{tr}$ is the number of seismic traces which are close enough to contain reflection data from that image point, and $f_K$ is the number of operations required per image point per trace. The sampling in each image dimension is of order 103, and the number of nearby seismic traces $N_{tr}$ is of order 107. The number of floating point operations per unit computation, $f_K$, is of order 10, yielding a total flop count of order 1017 or 3 Gflop-years; *i.e.,* 3 years of computation time on a 1-Gflop machine.

As expensive as this migration method is, it involves several limiting approximations which affect the quality of the result. The propagation times and amplitudes are derived from high-frequency (ray-theoretic) approximations and usually only account for a single propagation path. While there are well-known migration algorithms which properly handle finite-frequency effects and include multiple paths of propagation, their cost exceeds Kirchhoff migration by one or more orders of magnitude, making them prohibitively expensive on traditional supercomputers.

For seismic imaging to impact business decisions, the entire iterative process must take no longer than a few months. And since three or four iterations are usually performed, each iteration should take no more than a month. To cost effectively perform the several Gflop-years of computation for Kirchhoff migration in this timeframe requires a parallel supercomputer and an efficient parallel algorithm.

## 4. Approach and Methodology

In general, any algorithms that involves doing operations on a large set of data, where the problem can be broken down into smaller independent sub-problems can be described in MapReduce framework. More specifically, Chu et al in [8] describe that machine learning algorithms which can be written in a certain summation form can be parallelized easily on MapReduce. In the following we divided the PKTM algorithm in a summation form.

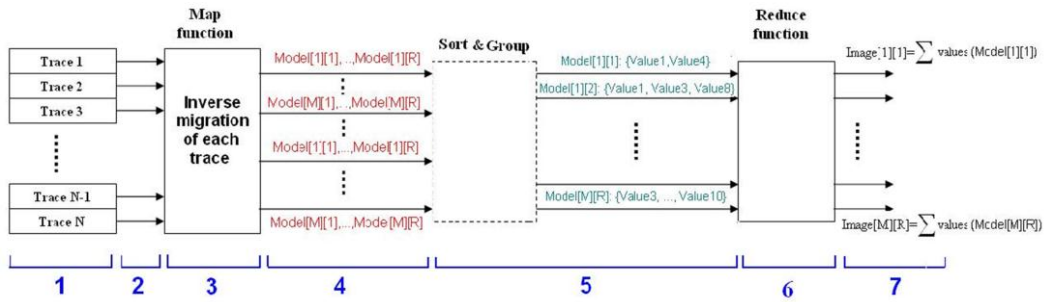

Figure 3. The proposed MapReduce framework for Inverse PKTM

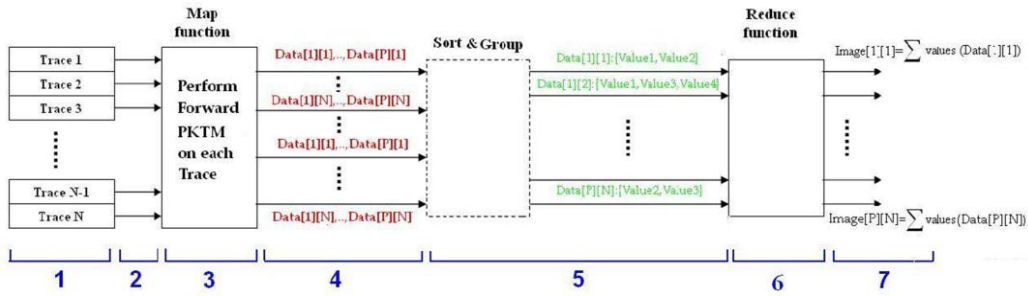

Figure 2. The proposed MapReduce framework for Forward PKTM.

*Forward PKTM*

Figure 2 shows the proposed MapReduce framework in [5] to develop Forward PKTM on cloud computing infrastructure:
1. The framework input is data traces (Trace 1, Trace 2,…, Trace N). If there are K sources and L receivers, then the number of traces will be N=K*L. Also, each trace includes P samples with sampling rate.
2. The primary key of the Map function is the index of a trace and the value is the sampling values of that trace.
3. Map function migrates each trace (by forward FKTM).
4. The size of the area is $M \times R$.
5. $Data[i][j], 1 \leq i \leq P \quad 1 \leq j \leq N$ indicates a point which its energy must be calculated. $Data[i][j]$ is the output of map function and it may be repeated several times (with different values) as each receiver is involved in several traces.
6. In Group/Sort section of the framework, the outputs of map functions are sorted based on the order of the $Data[i][j]$ indexes, e.g. $Data[1][1], Data[1][2],..., Data[P][N]$ .Also, the values of each element is the value of each occurrence of that element. For example in the figure 2, $Data[1][1]$ is repeated two times with values {value1, value2}.
7. In Reduce function, the values of all occurrences of each data ( $Data[i][j]$ ) are summed and formed the final image.

*Inverse PKTM*

Figure 3 described the proposed MapReduce framework for inverse PKTM. As can be seen the both forward and inverse MapReduce frameworks have the same data flow except:

- In stage (3), the traces are migrated by using inverse FKTM, which updates the velocity model.
- The final image is the update of velocity model.

## 4. CONCLUSION

In this report, two MapReduce frameworks have been studied to process forward and inverse PKTM algorithms on cloud computing. Most researches show that using PKTM is enough to build the seismic image for most cases. However, when the signal to noise is weak, it is recommended to use signal processing techniques (like filtering) to increase this ratio.